# Classification of networks-on-chip in the context of analysis of promising self-organizing routing algorithms

*Alina Rogulina*
*Student*
*HSE University*
*34 Tallinskaya Ulitsa, 123458, Moscow, Russian Federation*
*avrogulina@edu.hse.ru*

*Olga Volgina*
*Student*
*HSE University*
*34 Tallinskaya Ulitsa, 123458, Moscow, Russian Federation*
*oavolgina@edu.hse.ru*

*Aleksandr Romanov*
*PhD, Associate professor*
*HSE University*
*34 Tallinskaya Ulitsa, 123458, Moscow, Russian Federation*
*a.romanov@hse.ru*

*Andrej Sukhov*
*Doctor of Science, Professor*
*HSE University*
*34 Tallinskaya Ulitsa, 123458, Moscow, Russian Federation*
*amsuhkov@hse.ru*

**Abstract:** This paper contains a detailed analysis of the current state of the network-on-chip (NoC) research field, based on which the authors propose the new NoC classification that is more complete in comparison with previous ones. The state of the domain associated with wireless NoC is investigated, as the transition to these NoCs reduces latency. There is an assumption that routing algorithms from classical network theory may demonstrate high performance. So, in this article, the possibility of the usage of self-organizing algorithms in a wireless NoC is also provided. This approach has a lot of advantages described in the paper. The results of the research can be useful for developers and NoC manufacturers as specific recommendations, algorithms, programs, and models for the organization of the production and technological process.

**Key words:** network-on-chip, network-on-chip classification, self-organizing algorithms, wireless self-organizing networks.

## 1 Introduction

Nowadays, increasing clock speed almost does not influence the performance of processors. The main way of development focused on the creation of multiprocessor chips. At the same time, the number of coprocessors has been growing rapidly and it may reach hundreds of microprocessors with tens of thousands of cores in graphics chips. Nevertheless, it is impossible to use the classical bus architecture while increasing the number of cores [1]. Therefore, it is recommended distributing computing processes between computing resources and provide communication between them for a more effective work organization, which determines the development of the theory of network-on-chip (NoC) and demonstrates the relevance of research in this area.

In the classical theory of computer networks, enough algorithms, methods, and communication protocols have been developed that could become the basis of NoCs [2]. In this paper, it is planned to apply routing algorithms and methods that were developed for wireless self-organizing networks for NoCs. Since a structure of connections of cores is fixed, a network (although it will consist of hundreds or thousands of nodes) will remain static, which makes it advisable to use virtual coordinates that allow assigning a unique address to each node of the network (and such an assignment can be carried out in advance, at the stage of processor production).

Solving the problem of creating reliable and high-speed routing methods for NoCs means moving to a new stage of development of high-performance multicore computing systems. The created fundamental foundation should ensure the development of technologies for designing and manufacturing multicore chips for many years. The current point of view expressed by manufacturers and representatives of leading scientific schools is that NoCs are the main direction of development of computing systems in the foreseeable future.

## 2 Network-on-chip classification

Let us consider and summarize the knowledge about the classifications and types of NoCs that exist in the modern world. The most complete relative to the previously proposed solutions is shown in Fig. 1. The following paragraphs will discuss in detail the aspects presented in this classification.

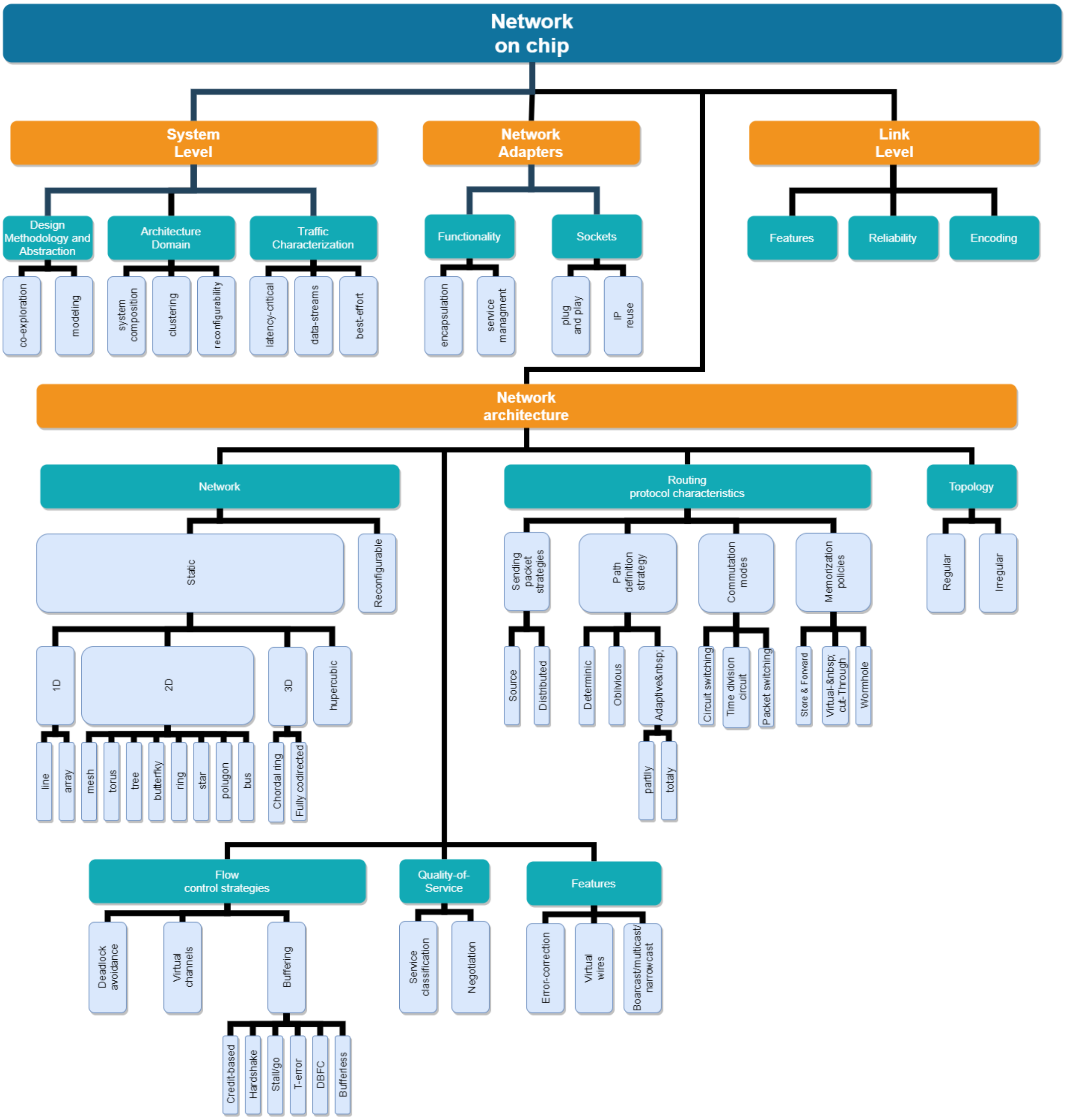

Figure 1 – NoCs' classification

### 2.1 NoC routing protocols

Network-on-Chip (NoC) is the most popular interconnection mechanism used for systems on chips that require flexibility, extensibility, and low power consumption. However, the performance of this mechanism is closely related to the routing algorithm used in NoC. The most important problems in the routing process are: deadlock, livelock, congestion, and faults. The classification of NoC routing protocols is proposed in [3] according to the problems they solve. There are two main groups: mono- and multi-objective routing protocols that solve one or more of the problems described above (Figure 2). This paper also shows that it is difficult to achieve solutions to all four aspects simultaneously using classical methods and, thus, emphasizes the strengths of multi-purpose approaches.

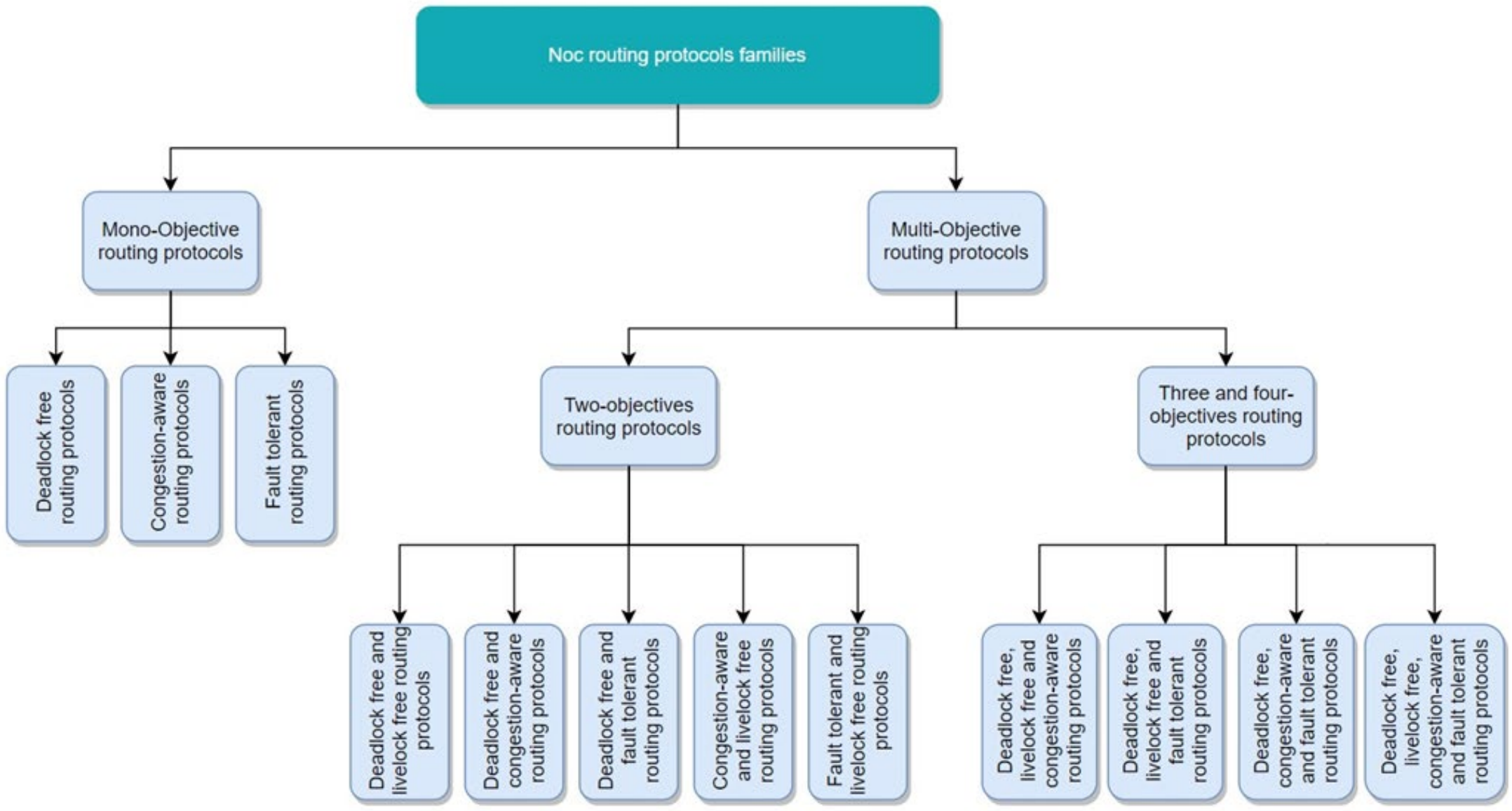

Figure 2 – NoC routing protocols' classification

We will also consider the works devoted to solutions of various types of malfunctions at different levels of NoCs. The article [4] presents the work performed on the methods of fault-tolerant mapping in NoCs. It also suggests the classification based on the approaches taken to recover from failures. The main categories of the proposed classification are methods based on the combination of mapping and routing, methods based on redundancy, and methods based on task reassignment (Fig. 3). In addition, this article compares the effectiveness of these methods.

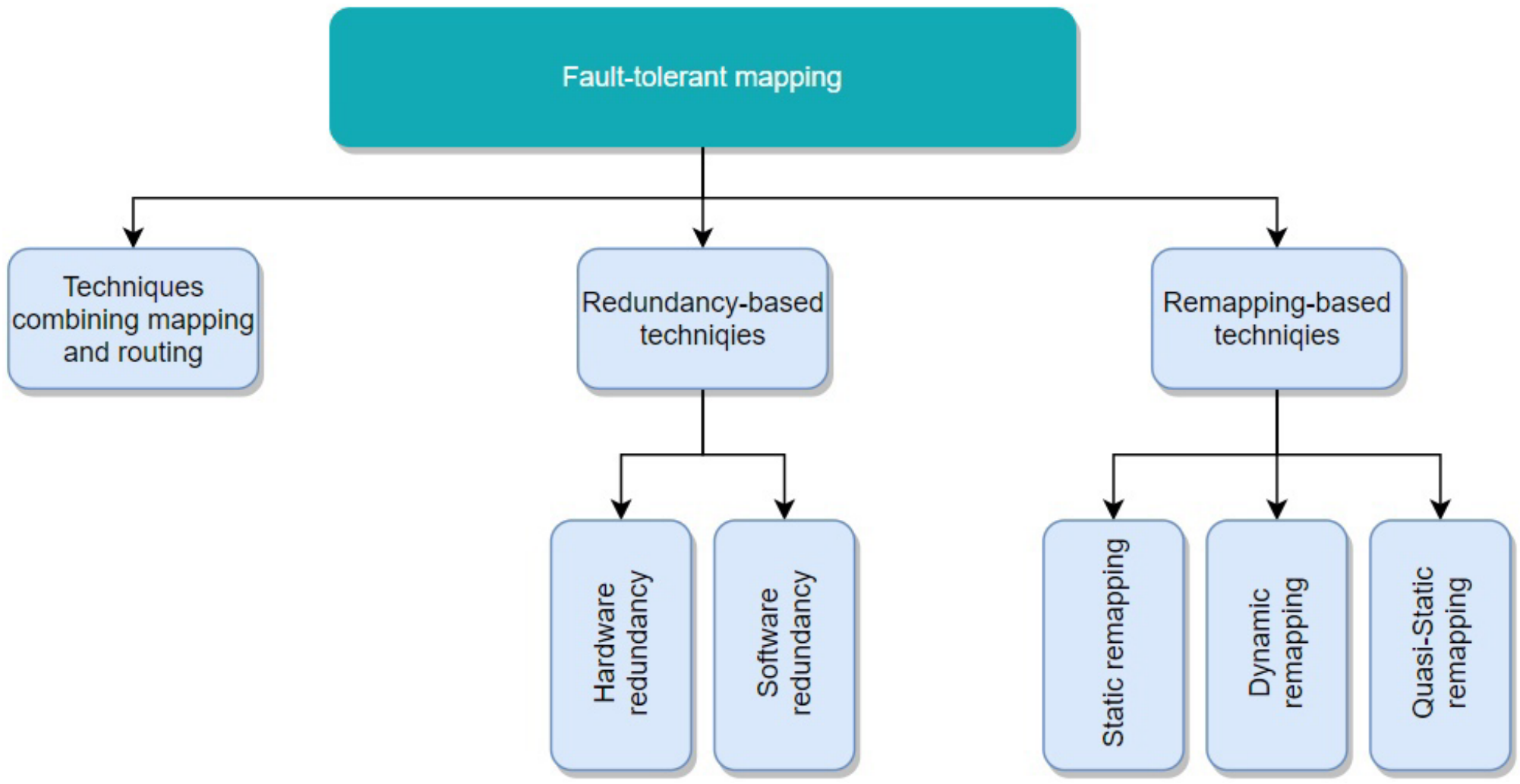

Figure 3 –NoC fault-tolerant mapping methods' classification

The article [5] presents the classification of routing algorithm methods according to various criteria, such as aging-aware, thermal-aware, congestion-aware, fault-aware, resilient, and energy efficient.

In the following sections we will consider congestions, flow control, and classification of NoCs topologies in more detail.

### 2.2 Congestions in NoC

NoC was introduced to improve the performance of chip multiprocessors and the execution of parallel programs. Although NoCs are known as modular and scalable interconnect infrastructures, they also have high latency and power consumption due to the communication between the cores. A potential solution to these problems may be a wireless NoC (WiNoC), which is able to provide high bandwidth and low latency using the unique features of wireless interconnects. However, wireless routers are prone to congestion due to the limited number of wireless channels on a chip and the sharing of these channels by all processing elements. In the study [6], a time-load balancing routing algorithm with congestion consideration (LTCA) is proposed to eliminate congestion in wireless routers and to balance the load distribution of traffic over wired and wireless networks. LTCA is a deadlock-free routing algorithm in which only a limited number of packets are allowed to use wireless channels. The required time for transmitting selected packets over wireless communication channels is measured considering the bandwidth of the wireless channels and the traffic load. The results of synthetic and real traffic modeling of 3-tuples showed the significant latency reduction, improving of the usage of wired and wireless communication lines and the probability of packet loss, as well as the increase in bandwidth.

Another adaptive routing algorithm, consistent with the traffic load, that solves the problem of congestion in wireless routers is proposed in [7]. The proposed algorithm selects the source-destination pairs with the largest wired transition distance for using wireless channels. The number of selected packets is determined based on the bandwidth of the wireless channel and the load of network traffic. The simulation results show the latency reduction up to 65 %, the usage of wired and wireless communication lines improvement by 16 % and the saturation bandwidth increase by about 11 %.

The article [8] proposes the algorithm for simultaneously ensuring the absence of a deadlock and a livelock – dynamic XY (DyXY). A new router architecture has developed to support the routing algorithm. In the same work, analytical models for DyXY routing for the two-dimensional grid architecture of NoC were designed, which coincided with the simulation results. It should be noted that DyXY routing can have higher performance compared to static XY routing and odd-even routing.

The existing interconnect networks use either unconscious or adaptive routing algorithms to determine the route along which a packet is sent to its destination. Despite the somewhat higher complexity of the implementation, adaptive routing has better fault tolerance characteristics, increases network bandwidth, and reduces latency during nonuniform or bursty traffic compared to oblivious policies. However, adaptive routing can degrade performance by disrupting any inherent global load balancer through greedy local solutions. To improve the load balance during routing adaptation, Regional Congestion Awareness (RCA) is proposed in [9] – a method for improving the balance of the global network. Instead of relying solely on local congestion information, RCA informs the routing policy about congestion in parts of the network outside of neighboring routers. The experiments in this article show that RCA meets or exceeds the performance of conventional adaptive routing in all the workloads studied with an average value of 16 % and a maximum delay reduction of 71 % in SPLASH-2 tests running on a 49-core CMP.

Another adaptive routing algorithm that can help minimize path congestion using load balancing is proposed in [10]. Conventional adaptive routing schemes use only channel information to determine the overload state. Due to the lack of switch-based information, it is difficult to identify the real state of congestion along the routing path only from channel-based information. Therefore, the authors of the article [10] suggest reconstructing the information about congestions of a path in order to show hidden spatial information about congestions and increase the efficiency of choosing a routing path. An adaptive routing scheme with consideration of path congestion (PCAR) is proposed, based on the following methods: 1) a selection strategy considering path congestion, which simultaneously considers switch congestion and channel congestion, and 2) a competition prediction method that uses the rate of change of the buffer level to predict possible switch competition. The experimental results show that the proposed PCAR scheme allows achieving high saturation throughput with an improvement of 15.4 %-48.7 % compared to the existing routing schemes. The PCAR method also includes the VLSI architecture, which has a higher area efficiency with an improvement of 16 %-35.7 % compared to the other router designs.

### 2.3 Flow control in NoC

Flow control mechanisms in the NoC architecture are crucial for fast packet propagation over the network and low idling of network resources. Buffer management and allocation are the main tasks of each flow control scheme. The alternative schemes, such as SHALL/GO, T-Error, ACK/NACK, for allocating buffer and channel bandwidth in the presence of switch-to-switch pipeline communication channels are considered in [11]. These protocols provide different degrees of fault tolerance support, which leads to different power tradeoffs. The analysis presented in this paper is aimed at determining the overhead costs of support for a given course of work in error-free environments (they are typical operating modes).

Since the bandwidth offered by NoC can be effectively used only if there are effective flow control algorithms and the existing algorithms today either rely only on local information or have large communication costs and unpredictable delays the authors of [12] proposed a predictive flow control mechanism with a closed loop. The following changes were also made. Firstly, models of traffic sources and routers specifically designed for NoCs were developed. Secondly, these models were then used to predict cases of possible network congestion. Finally, based on this information the authors proposed the scheme for controlling the packet injection rate into traffic sources to regulate the total number of packets in the network. Estimates using real and synthetic traffic models have shown that the proposed algorithm provides better performance compared to traditional algorithms for controlling the flow from switch to switch.

The flow control of a buffered channel and a non-buffered channel with packet switching are compared in [13]. The evaluation proposed in this article includes optimization for both schemes: buffered networks used custom SRAM-based buffers and empty buffer bypass to increase energy efficiency, while buffer-free networks used the new routing scheme that provided a 5 % reduction in average latency. The results showed that if technological limitations do not lead to excessively expensive buffers, the performance and increased complexity of managing the deviating flow (i.e., without buffering) outweigh its potential benefits: buffer-free designs are only slightly (up to 1.5 %) more energy efficient at very low loads, and buffer networks provide less latency and higher throughput per unit of power in most cases.

The new flow control mechanism that provides flexibility for routing each packet flit independently, more effectively reducing network congestion in the routing of NoCs based on a virtual channel is proposed in [14]. In this case, 100 % buffer utilization is achieved by eliminating the overhead of virtual channel arbitration. The modified design of the router has an 80 % smaller area and consumes 473 times less energy compared to the conventional design of the router. With uniform random traffic, there is a decrease in packet latency by 35-78 %, 27-68 % and 45-62 % for the following architectures, respectively: mesh, torus, and flattened butterfly topology. The average packet drop rate is reduced by at least 28 % in the above topologies.

### 2.4 NoC topologies

System-on-chip applications require additional performance due to their complexity and heterogeneity. Consequently, the need for more efficient communication mechanisms is constantly growing. That is why the NoC paradigm was created since computer networks.

Network-on-chip includes both topology and design methodology. The topology of the NoC is a network of switches, resources are placed in the slots formed by the switches. The resource can be a processor core, memory, FPGA, custom hardware block, or other intellectual property block that fits into an available slot and conforms to the NoC interface. The NoC architecture is, in fact, a communication infrastructure on a chip, consisting of a physical layer, a data transmission channel layer, and a network layer of the OSI protocol stack [15]. The NoC design methodology consists of two stages. First, a specific architecture is inferred from the general NoC template. Second, the application is mapped to a specific architecture to form a specific product.

Networks-on-chip were built for the following parameters: bandwidth scalability compared to traditional bus architectures; energy efficiency and reliability; possibility of reuse [16].

The article [17] discusses the classical topologies for constructing NoC and their main advantages and disadvantages. Also, the algorithm for finding optimal topologies was proposed and implemented in accordance with the restrictions on the diameter and maximum degree of vertices with optimization for the number of connections and average distance. Optimal topologies are synthesized for vertices from 6 to 10.

The work [18] presents the implementation of several routing algorithms of the type intended for use in networks with circulant topology such as $C_{(N;\,1,s_2,s_3)}$ to find the shortest routes between any two network nodes. The algorithms were also tested on various sets of optimal speed circulants and compared in terms of efficiency and memory resources.

The authors of the article [19] conducted studies of NoCs with 2D-mesh and 2D-torus-mesh topologies for multi-core processors with the Elbrus architecture. In the course of work, it is necessary to use the topology of 8- and 12-core processors, which provides a large throughput of the 2D mesh, while for a high-performance 16-core processor, it is necessary to use the 2D-torus-mesh topology.

The article [20] presents the NoC architecture, which allows reconfiguring the network topology. In this way, the architecture allows for the generalized system-on-chip platform in which the topology can be configured for the applications currently running on the chip, including long links and direct links between IP blocks. Configurability is inserted as a layer between routers and communications systems. The topology is configured using energy efficient topological switches based on physical circuit switching as shown in the FPGA. The article introduces the ReNoC (Reconfigurable NoC) architecture and simulations show the 56 % reduction in power consumption over static 2D mesh topology (2D mesh topology).

Thanks to the inclusion of the third dimension, several interesting topologies appear in networks-on-chip [21]. The speed and power consumption of 3D NoC are comparable to those of 2D NoC. In calculating the speed and power consumption of these three-dimensional structures physical constraints are used, such as the maximum number of planes that can be vertically complex and the asymmetry between horizontal and vertical network communication systems. Also, in this work the analytical model of the zero-load delay of each network, considering the influence of the topology on the three-dimensional RNC, was proposed. The trade-off between the number of nodes used in the third dimension, reducing the average number of hops, traversable packets, physical planes, integration of network functional blocks that reduce the length of the communication channel, is estimated for both the delay and the power consumption of the network. Performance improvements were seen by 40 % and 36 % and power consumption decreased by 62 % and 58 % for 3D NoCs compared to traditional 2D NoC topology for network sizes of N = 128 and N = 256 nodes respectively.

With the trend towards more cores in chip multiprocessors, the on-chip interconnection between cores must scale efficiently. The article [22] proposes the usage of high radius networks-on-chip interconnect networks and describes how a flattened butterfly topology can be mapped to networks-on-chip. By using high-range routers to reduce the diameter of the network, the flattened butterfly provides lower latency and lower power consumption than conventional on-chip topologies. In addition, using a two-dimensional flat VLSI circuit, the flattened butterfly-on-chip can use bypass channels so that non-minimal routing can be used with minimal impact on latency and power consumption. This paper evaluates the flattened butterfly and compares it to alternative on-chip topologies using synthetic traffic and routing patterns and demonstrates that the flattened butterfly can increase bandwidth by up to 50 % over a concentrated mesh and reduce latency by 28 % while reducing energy consumption by 38 % compared to a mesh network.

The authors of [23] proposed the idea of disintegration by splitting a large chip into many smaller chips and using interposer-based (2.5D) integration to connect smaller chips. This method can improve performance, but as the number of small chips increases, communication between chips becomes a performance bottleneck. Therefore, this article proposes the new network topology called "ClusCross" to improve network performance for multi-core interconnects in silicon interposer systems. The key idea is to treat each small chip as a cluster and use cross-cluster long links to increase the bisection width and decrease the average hop count without increasing the number of ports in the routers. Synthetic motion models and real-world applications are simulated on an accurate cycle simulator. In the course of the work, a decrease in network latency and an increase in saturation throughput were found compared to previously proposed topologies. Two versions of the ClusCross topology are evaluated: one of the ClusCross versions has an average latency reduction for coherent traffic of 10 % compared to the modern network-on interposer topology that is not aligned with ButterDonut; another version of ClusCross has a 7 % and 10 % reduction in power consumption compared to the FoldedTorus and ButterDonut topologies, respectively.

Two types of parameters are required to design RNC routing algorithms: those related to the RNC architecture, and those that are defined by the routing protocol itself. The architecture parameters allow to define the network topology, which can be regular or irregular [3]. Irregular, in turn, can be specialized or quasi-optional. The above parameters also define the type of networks (static or dynamic) and the network topology, which can be regular/2D/3D or hierarchical. The previously described classification of the network-on-chip topology is clearly shown in Fig. 4.

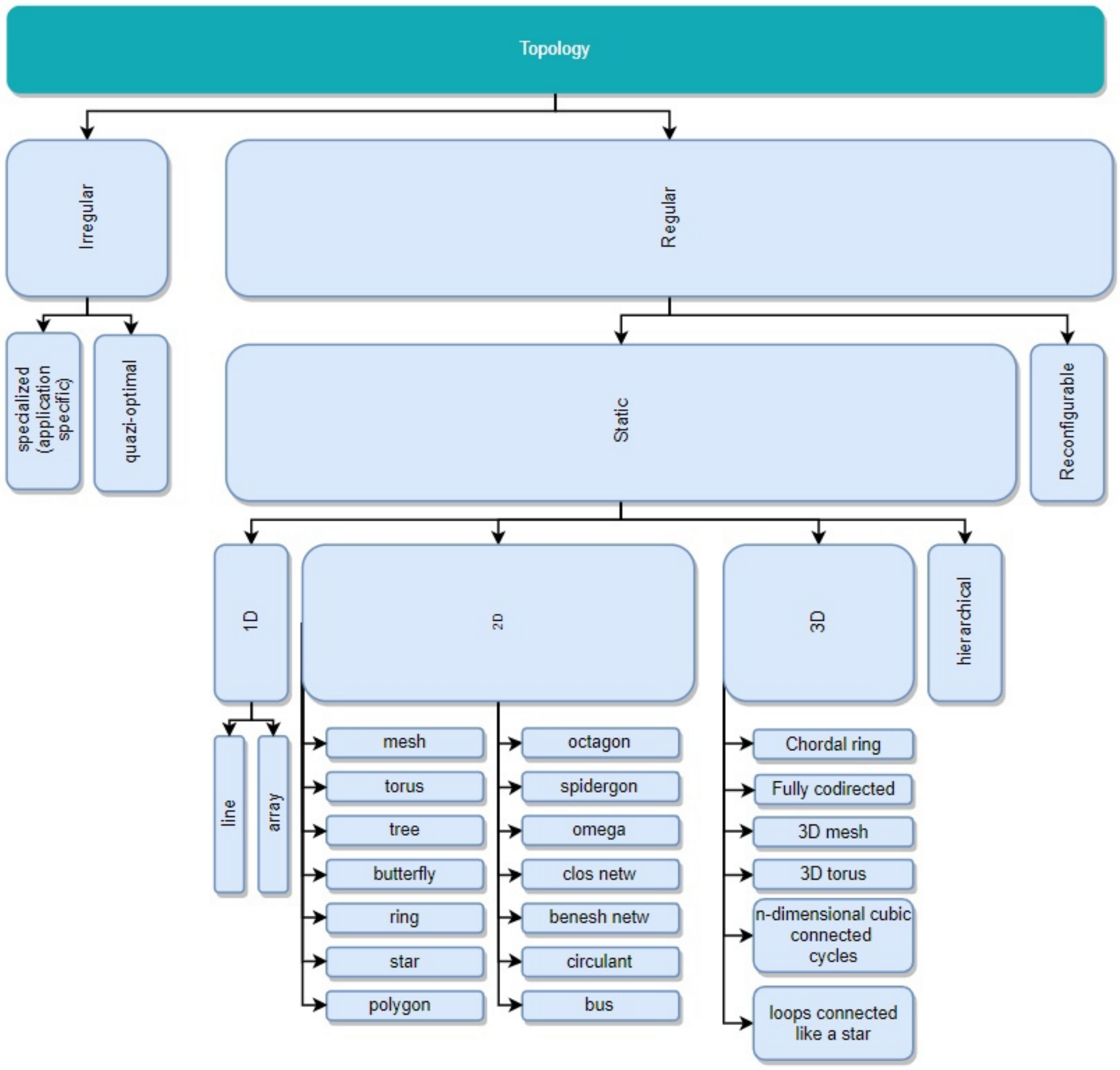

Figure 4 –NoC architectures' classification

### 2.5 Wireless networks-on-chip

The problem of signal delays and desynchronization often arises during designing systems-on-chip. Various methods of organizing network structures on a chip, which theoretically allow minimizing delays and losses, have drawbacks and are difficult to implement. The most promising way to solve this problem is the paradigm of wireless networks-on-chip, which allows to bypass the limitations of classical networks, as well as to provide communication between nanoscale components of microcircuits and the macro level [24].

In-chip communication is primarily based on the network-on-chip design paradigm in modern multiprocessor chips and multiprocessor systems-on-chip. However, it should be highlighted that conventional NoC architectures will not be able to meet the performance, power, and reliability requirements of the next generation of multicore architectures [25]. More recently, on-chip communication technologies such as wireless RNCs have been proposed as possible solutions to overcome the scalability limitations of traditional multi-node RNC architectures.

Wireless system-on-chip is a promising candidate for solving the performance and efficiency issues that arise when scaling existing NoC technologies onto multi-core processors. Existing wireless RNC architectures typically use millimeter-wave antennas without significant directivity gain, as well as a token transfer protocol for accessing a common wireless environment. The antennas provide built-in wireless communications for data transmission, minimizing transmission delay, and energy dissipation [26]. However, there is a limitation on the achievable performance advantage, since only one wireless pair can communicate at a time, otherwise the wireless communication may cause interference.

Wireless RNC can handle WAN and broadcast traffic with ultra-low latency even on microcircuits with thousands of cores, thereby acting as a natural complement to traditional and bandwidth oriented wired RNC [27].

## 3 Investigation of self-organizing routing algorithms in networks-on-chip

Learning a routing method is a complex task that requires several basic and exploratory research. First, it is required to define a routing algorithm that provides for the possibility of communication between remote computing cores with a significant increase in their number. In this case, it is necessary to introduce the concept of the distance between the nodes of the network so that when the graph of the computational problem is projected onto the network topology, communication is carried out between the closest nodes. This approach reduces communication losses in computations. The essence of the latest research is to study the application of routing algorithms used in self-organizing wireless networks in a new area – networks-on-chip. Since the structure of the cores' connections is fixed, the network will remain static, which makes it expedient to use virtual coordinates by analogy with wireless self-organizing networks.

The fact that the structure of links inside the chip cannot be changed makes it possible to introduce an address system even at the stage of chip production. The address system specified once will be used throughout the entire operation of the chip. This fact is very important, it allows avoiding unnecessary waste of computing resources and saving power consumption during the operation of the multiprocessor system-on-chip.

One of the main requirements for modern routing algorithms is the absence of routing tables. For example, widely used routing methods, for example, in TCP / IP networks, face big problems associated with the growth of routing tables. Their growth is proportional to the increase in the number of nodes in the network. Keeping the complete routing table in RAM is challenging. If one abandons routing tables in the process of building a route, then information about the route must be contained in the address system. For example, such a method was proposed in [28]. This method combined the introduction of a hierarchical addressing system and the principle of greedy routing. However, it has an important disadvantage: it is the congestion of the upper channels of the hierarchy, since most of the traffic is forced to use these channels. To solve this problem, it was proposed to use several alternative centers of the hierarchy. These two approaches (hierarchical routing with multiple routing centers and an address system that does not require routing due to the peculiarities of its construction) seem to us the most promising direction for finding new routing algorithms in NoCs.

The routing algorithm should be characterized by the strategy that is used to send packets if they are originating or distributed; its strategy for determining the path, which can be deterministic, adaptive (partially or completely), that is, determined in the process of solving a problem based on the accumulation of new information, or non-adaptive; its switched mode, which can be packet switched, circuit switched, or a hybrid mode called time division. Finally, the routing protocol must use one of the following policies for storing flats in router buffers: Store and Forward, Virtual-cut-Through, or Wormhole [3].

The previously described classification of routing protocols in networks-on-chip is clearly shown in Fig. 5.

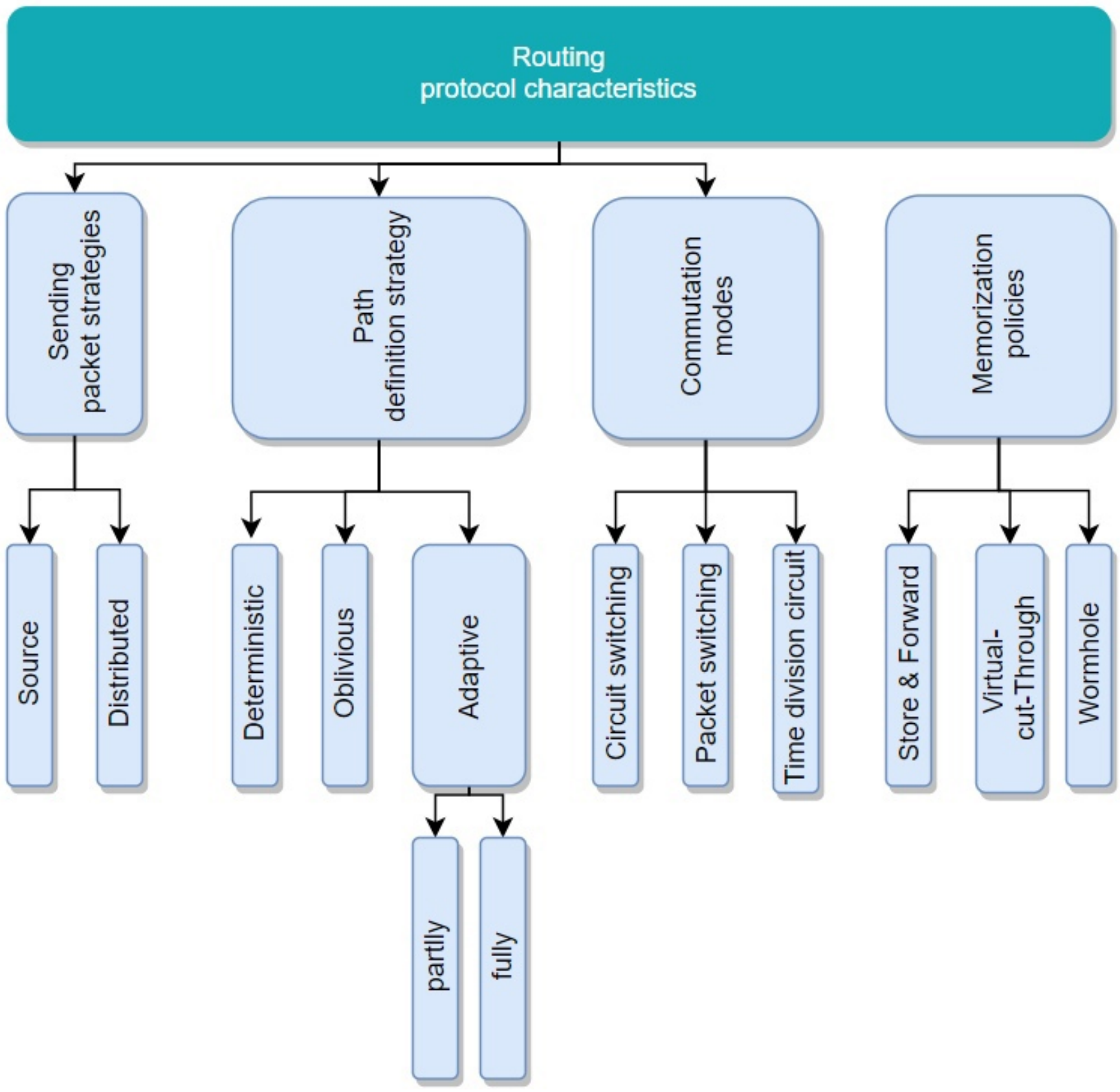

Figure 5 – NoC routing protocols' classification

The distance between the nodes of the network is the basis for the method of greedy forwarding, which allows finding the shortest paths between the cores. Based on virtual coordinates, the concept of a vector between final nodes of the route is introduced and the principle of greedy advance is formulated. This principle assumes the choice of the next node of the route as one of the neighboring nodes of the current node. The definition of the proposed coordinate system and the principle of greedy advancement are studied since high-level models.

At this point, the research has already presented the greedy topology-independent protocol for routing in networks-on-chip. The proposed routing algorithm admits any number of permanent failures and as proved, situations where unsolvable network blockages occur. A specialized version of the algorithm is optimized for two-dimensional mesh networks, both classical wired and wireless [29]. The adaptability and minimality of several variants of this algorithm are analyzed using graphical modeling.

In [30] and [31], the neighborhood method was described that builds a route using two broadcast requests. At the first stage, neighborhoods are labeled relative to the starting node of the route. As soon as the markup reaches the end node, the second stage begins. At this stage, a route is constructed as a sequence of nodes lying at the intersection of the first neighborhood of the current node and the corresponding neighborhood of the initial node. In this case, the process of building a route begins from the final node. Thus, it is possible to build a set of all possible routes.

## 4 Evaluation

We performed an exhaustive classification of NoCs, highlighted the macro levels of classification, such as System Level, Network Adapters Level, Link Level and Network Architecture Level, and analyzed them at the levels below, down to the levels of local features, such as topology type or flow-control strategy. We analyzed special cases of NoC implementations and specific decisions made at various levels of the classification. The results of the analysis allow us to say that NoCs (if we consider them as a transmission medium and switching data streams) have many similarities with wireless networks, such as the same Link Level, Network Architecture Level, etc.; and decisions made at the quality-of-service or routing protocol level affect the functioning of the network too.

Of course (due to the physical differences in the principles of data transmission, since in NoC data is transmitted directly in the chip; and in wireless networks, the transmission environment is radio waves and switching devices) there are certain differences. For example, in NoCs there is no urgent need to ensure the immunity of the transmission protocol (or it is done much more simply by simple parity control) and its encryption from unauthorized access. But some of the seemingly differences, with a deeper analysis, can be eliminated or replaced by other abstractions. So, wireless networks are known for their self-organization methods when the transmission protocol and routing algorithm are configured for a dynamically changing environment; individual devices that make up the network can be disconnected or connected, and their communication quality can decrease, etc. In the context of NoCs, it is impossible at first glance to say that their topology of connections between nodes is dynamically changing. But at the same time, considering the problems that arise in NoCs (deadlocks, livelocks, congestion, starvation, network faults), we can apply the method of simplification and reduce them all to network faults [3], i. e. temporary accidental disconnection of nodes or entire network segments. From this point of view, NoC no longer looks static, which means that self-organizing routing methods are quite applicable here. Moreover, since it is initially known how many nodes there will be in NoC and its initial topology, it becomes possible to pre-map the network by giving each node such coordinates, on the basis of which routing can be carried out (taking into account the fact that some nodes may become temporarily unavailable).

The proposed method of reflection and borrowing ideas from related fields of engineering can be extended to other areas of NoC design, such as the use of the reduced OSI model (from TCP / IP networks) for organizing a network protocol or methods of multiplexing data transmission lines between routers (from the area of peripheral communication interfaces, such as $I^2C$, for example) to reduce hardware costs for connections at the expense of lower transfer rates.

## 5 Conclusion

Based on the analysis carried out, the classification of topics for researching networks-on-chip was compiled. The current state of research on this issue allows to identify several key areas of development. The NoC architecture, the parameters of which characterize the network topology, determines the method for designing complex integrated circuits to reduce hardware resources and power consumption of the network-on-chip, as well as design time. Network performance is calculated using various factors, but bandwidth is the most important metric. To overcome the bottleneck in system performance and reliability caused by on-chip interconnects in networks-on-chip, the mechanism is being actively developed that takes advantage of wireless communication. The latest research in this area is looking at the application of the routing algorithms used in self-organizing wireless networks to the new area – networks-on-chip. As a result of the analysis, the main criteria for the classification of network-on-chip studies were identified.

Also, during the analysis of the existing literature, it was found that the usage of self-organizing algorithms in networks-on-chip is a rather unique method of routing in NoC. These ideas were recently proposed, and published experiments have shown that the use of self-organizing algorithms in NoC is possible. During further research, it is planned to test some of the self-organizing algorithms [3]. It should be noted that self-organizing wireless networks and NoCs have both a number of similarities and differences. The use of NoC features, such as a static nature (when the structure of the network, including the set of nearest neighbors, remains unchanged), the simpler organization of routers, communication channels, and the absence of the need to protect the network from malicious actions, allow us to assert the possibility of using all the advantages of self-organizing algorithms (especially if consider a NoC as a dynamic system with node failures for a short time), but without significantly complicating routing in the NoC.